\documentstyle[11pt,psfig]{article}

\def\chem#1{${}^{#1}$}
\def\G{{\cal G}}

\def\Macc{\dot{\mathrm M}_{\mathrm acc}}
\def\lacc{L$_{\mathrm acc}$}

\def\msun{$\mathrm{M_{\odot}}$}
\def\Msun{\mathrm{M_{\odot}}}

\def\Rsun{\mathrm{R_{\odot}}}

\def\Lsun{\mathrm{L_{\odot}}}
\def\myr{M$_\odot\,$yr$^{-1}$}
\def\Myr{\mathrm M_\odot\,yr^{-1}}
\def\la{\mathrel{\hbox{\rlap{\hbox{\lower4pt\hbox{$\sim$}}}\hbox{$<$}}}}
\def\ga{\mathrel{\hbox{\rlap{\hbox{\lower4pt\hbox{$\sim$}}}\hbox{$>$}}}}

\begin{document}

\begin{center}
{\LARGE \bf The Swallowing of Planets by Giant Stars \\ }
\vspace{0.5cm}
Lionel Siess\footnote{Space Telescope Science Institute, 3700 San Martin
Drive, Baltimore, MD 21218, USA}$^,$\footnote{Laboratoire d'Astrophysique de
Grenoble, BP 53X, 38041 Grenoble Cedex 9, France } and Mario
Livio$^{\tiny 1}$

\end{center}

\begin{abstract}
We present simulations of the accretion of a massive planet or brown dwarf
by an AGB star. In our scenario, close planets will be engulfed by the
star, spiral-in and be dissipated in the ``accretion region'' located at
the bottom of the convective envelope of the star. The deposition of mass
and chemical elements in this region will release a large amount of energy
that will produce a significant expansion of the star. For high accretion
rates, hot bottom burning is also activated. Finally, we present some
observational signatures of the accretion of a planet/brown dwarf and we
propose that this process may be responsible for the IR excess and high
lithium abundance observed in 4-8\% of single G and K giants.
\end{abstract}

\section{Introduction}

The discovery of planets around several main sequence solar type stars has
unexpectedly revealed that a large fraction of these ``hot Jupiters'' orbit
very close to the central star, at less than 1 AU (e.g. Mayor \& Queloz
1997). A major theoretical problem is to account for the location of these
planets but it also raises the question of their fate.  Indeed, close
planets like the companion of 51 Peg will ultimately be engulfed in the
star's envelope. The general scenario for the swallowing of a planet is
thus as follows~: planets in close orbits will be engulfed in the envelope
as the star leaves the main sequence and becomes a giant. Depending on the
orbital separation, the swallowing can take place along the red giant
branch or during the AGB phase. Then, due to viscous and tidal forces, the
planet will spiral-in and be dissipated near the center. The study of the
swallowing of planets has been previously investigated by several
researchers. Notably, Livio and Soker (1984) found that there exists a
critical mass below which the planet evaporates or collides with the core.
For the particular AGB star model used, they found a critical mass of the
order of $\sim 0.02$\msun, but these results must be taken with caution
since the physical processes involved in these calculations were treated
only approximately. Moreover, the outcome of the spiraling-in certainly
depends on the structure and evolutionary status of the star. 

In the next section we present the physical parameters of the accretion
process and our initial model. Then in $\S$\ref{result}, we present the
results of our simulations and finally, in $\S$\ref{conclusion}, we
conclude and describe the possible observational signatures of this event.

\section{Physical approach}

To determine the locus where the planet/brown dwarf will be dissipated, we
have evaluated different physical quantities. First, we have estimated the
Virial temperature of the brown dwarf which represents the temperature
above which the thermal pressure of the gas overwhelms the gravitational
binding energy. Its expression is given by
\begin{equation}
\mathrm{ \displaystyle T_{\mathrm{virial}} \sim 2.4\,10^6
\Bigg(\frac{M_{bd}}{0.01\Msun}\Bigg)
\Bigg(\frac{R_{bd}}{0.1\Rsun}\Bigg)^{-1}\ {\mathrm K}} \  ,
\end{equation}
where M$_{\mathrm bd}$ and R$_{\mathrm bd}$ are the mass and radius of the
brown dwarf. Then, we have estimated the locus in the star where the
elongation stress due to tidal effects becomes of the order of the central
pressure of the brown dwarf. Assuming the brown dwarf can be approximated by a
polytrope of index $n=1.5$, we found a critical radius
\begin{equation}
\mathrm{ R_{\mathrm{tidal}} \sim 0.3\,\Bigg(\frac{M_{bd}}{0.01\Msun}
\Bigg)^{-1/3} \Bigg(\frac{R_{bd}}{0.1\Rsun}\Bigg)\Bigg(
\frac{M_{*}}{\Msun}\Bigg)^{1/3} \Rsun}\ , 
\end{equation}
below which the brown dwarf is totally disrupted. Temperatures of the order
of a few million Kelvin and critical radius of about a solar radius are
typically encountered at the base of the convective envelope of giant
stars. Thus, it is likely that the brown dwarf will be dissipated at the
base of the convective envelope, in this so-called accretion region. To
evaluate the mass accretion rate, we have calculated the decay timescale of
the brown dwarf's orbit in the vicinity of the dissipation region. This
provides an estimate of the rate at which mass is flowing into the accretion
region and for typical parameters we found $\mathrm{\Macc \sim
10^{-5}-10^{-4}}$\,\myr.\\ Finally, we modified the chemical abundances in
the shells where the mass is accreted and we assumed that deuterium is not
present in the accreted matter. Our initial model is an AGB star of initial
3\msun and during the accretion process the matter is deposited from the
top of the H burning shell (located close to the base of the convective
envelope) up to the surface. Our simulations indicate that the profile of
accreted matter does not affect substantially the results. The important
point is how deep the mass is deposited and in the following discussion, we
identify the accretion region as the region located close to the bottom of
the convective envelope where the gravitational potential is very high.

\section{The evolution of the structure}
\label{result}

The star has to respond to the deposition of gravitational energy,
associated with the accretion luminosity 
\begin{equation}
\mathrm{ L_{acc} = \frac{\G M \Macc}{R} \approx 5650 \,
\Biggl(\frac{M}{0.54\,\Msun}\Biggr) \Biggl(\frac{R}{0.3\,\Rsun}\Biggr)^{-1}
\Biggl(\frac{\Macc}{10^{-4}\Myr}\Biggr)\,\Lsun}\ .
\end{equation}
Depending on the accretion rate, \lacc\ can represent a large fraction of the
stellar luminosity and thus can affect the structure significantly. We will
first analyze the effects of relatively high accretion rates.

In the case of high accretion rates ($\mathrm{\Macc = 10^{-4}}$\myr), the
release of potential energy produces a large expansion of the star. The
luminosity profile increases rapidly in the accretion region and presents a
bump that will progressively move outward and dissipate. Once a thermal
equilibrium is attained, the luminosity bump disappears and the star
contracts. The decrease in temperature due to the initial expansion is
responsible for an increase in the opacity. Rapidly, the radiative gradient
($\nabla_{\mathrm rad} \propto \kappa$L) becomes larger than the adiabatic
gradient and convection sets in. The convective envelope deepens and stops
when it reaches the H burning shell (HBS) : this marks the onset of hot
bottom burning and direct heating of the envelope. The sudden arrival of
\chem{12}C from the envelope temporarily increases the nuclear energy
production rate but the further advance of the HBS into the region that has
been expanded (and cooled down) will finally lead to a strong decrease in
the nuclear luminosity (Fig. 1). Thereafter, the accretion region acts as a
source shell, providing most of the stellar luminosity.

\begin{figure}[t]
\psfig{file=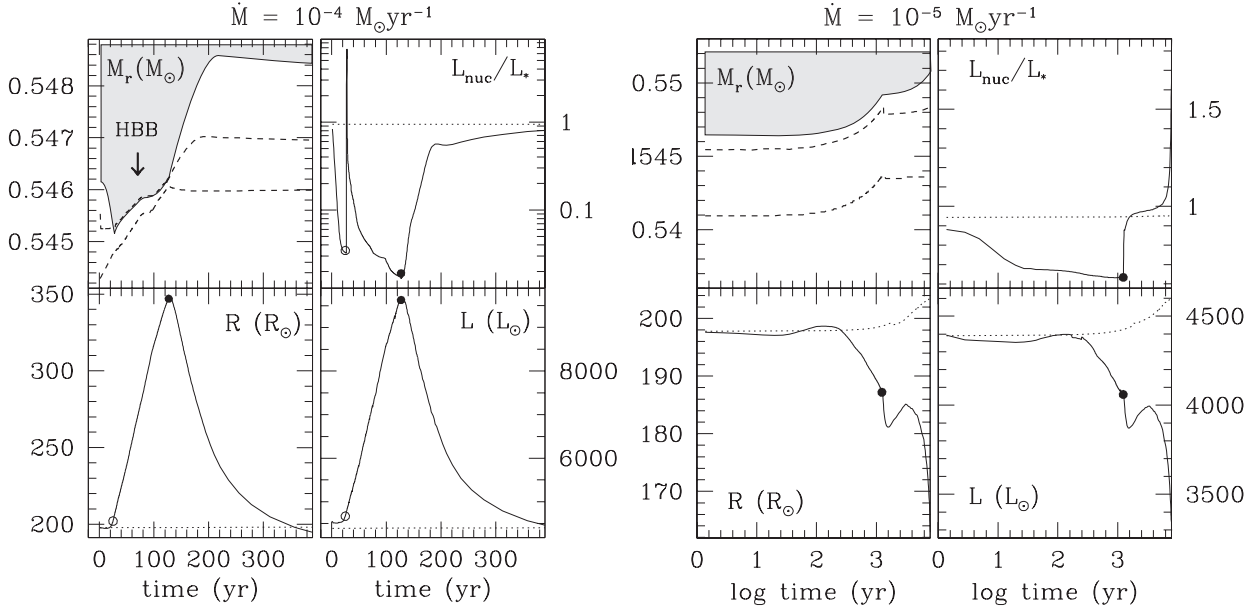,width=\textwidth}
\caption{Effects of the accretion of a 10 M$_{\mathrm{Jup}}$ planet. The
evolution of the boundaries of the HBS and convective envelope is depicted
in the upper left panels. The dark area indicates the convective envelope
and the dashed lines delineate the H burning shell ($\varepsilon_{\mathrm
nuc} > 5$ erg\,g${}^{-1}$s${}^{-1}$). In the other panels, the solid lines
correspond to the models with accretion while the dashed lines refer to
standard models (without accretion). L$_{\mathrm nuc}$ represents the total
nuclear luminosity, L and R are the stellar luminosity and radius,
respectively. Open and filled dots mark the beginning of the hot bottom
burning and the end of the accretion process, respectively.}
\end{figure}
For relatively lower accretion rates ($\mathrm{\Macc = 10^{-5}}$\myr), the
evolution is different. In this case, the release of accretion energy is
too small to produce a significant expansion of the star. Instead, due to
the increase in the core mass, the gravitational pull is reinforced and the
star contracts. Throughout the structure, the temperature increases and
globally the evolution is accelerated. Note that the nuclear energy
production rate keeps a constant value, around $\sim 70\%$ of the stellar
luminosity, the remaining $\sim 30$\% being provided by the accretion
process. In conclusion, for (relatively) low accretion rates, due to the
low value of \lacc, it is mainly the mechanical effects of mass deposition
and gravitational attraction that prevail.

To summarize, the stellar response to the accretion of a planet/brown dwarf
depends on the relative magnitude of the thermal adjustment timescale of
the star $\tau_{\mathrm \tiny KH} \simeq \G\mathrm M^2/RL$ to the accretion
time scale $\tau_{\mathrm acc} \sim\mathrm M/\Macc$. The star will expand
until $\tau_{\mathrm {\tiny KH}} \simeq \tau_{\mathrm acc}$ after which a
thermal equilibrium is reached and contraction resumes.  The effects are
thus dependent on the evolutionary status of the star ($\tau_{\mathrm
{\tiny KH}}$) and on the accretion rates ($\tau_{\mathrm acc}$).

\section{Conclusion}
\label{conclusion}

The swallowing of a giant planet/brown dwarf can substantially affect the
structure of giant stars and thus can have important observational
signatures. \\ During the spiraling-in process, the orbital angular
momentum of the planet is progressively imparted to the envelope. This will
certainly increase the rotation rate of the star and may produce very fast
rotators such as FK Comae-type stars. The deposition of angular momentum in
the accretion region may also induce some shear at the base of the
convective region. This, in turn, could generate strong magnetic fields by
a dynamo effect and lead perhaps to an enhanced X-ray emission due to the
confinement of the chromospheric plasma by the generated magnetic
field. The modifications to the structure and the increase in radius and
luminosity can also substantially increase the mass loss rate according to
Reimer's law ($\mathrm{\dot M_{loss} \propto LR/M}$). We can therefore
expect the formation of shells of mass of the order of $\sim10^{-5} -
10^{-3}$\msun, corresponding to individual accretion events. Finally, the
deposition of matter with a different chemical composition from that of the
envelope may affect the surface chemical abundances and lead to stellar
metallicity enhancements (see also Sandquist et al. 1998). Assuming a solar
composition for the planet/brown dwarf, the most important effects concern
\chem{7}Li whose abundance can be significantly increased. The changes in
the surface abundances are mainly due to dilution effects, and thus are
strongly dependent on the giant's envelope mass and on the planet/brown
dwarf mass.

Recently, de la Reza et al. (1996, 1997) noticed that most of the Li rich
giants, which account for $\sim 4-8$\% of the G and K giants (Brown et
al. 1989), also exhibit an IR excess compatible with the emission from a
circumstellar shell. These authors then suggested that Li is produced via
the Cameron-Fowler mechanism (1971), in which \chem{7}Be is dredged up from
the center to the envelope where it decays through the \chem{7}Be($e^-,
\nu$)\chem{7}Li reaction. They also assumed that this transport process is
accompanied by the ejection of matter, but they did not specify the
mechanism responsible for the enhanced mass loss. Instead, we propose that
the strong correlation between high Li abundance and IR excess could be
consistently explained in terms of the accretion of a massive planet or brown
dwarf by a giant stars. Our model can indeed account for both the mass
ejection and the Li enrichment. Interestingly, the $\sim 4-8$\% Li rich G and K
giants that show an IR excess are all single stars.

\section*{Acknowledgments}

LS acknowledges support from the Director's Discretionary Research Fond at
STScI and thanks the institute for its hospitality. This work has been
supported in part by NASA grants NAGW-2678, G005.52200 and G005.44000.

\end{document}